\documentclass[12pt]{article}
\usepackage[export]{adjustbox}
\usepackage{graphicx}
\usepackage{graphics}
\usepackage{epsfig}
\usepackage{amssymb}
\usepackage{amsmath}
\usepackage{color}
\usepackage{textcomp}
\usepackage{latexsym}
\usepackage{physics}

\usepackage[margin=2.5cm,footskip=0.6cm]{geometry}
\def\bea{\begin{eqnarray}}
\def\eea{\end{eqnarray}}
\def\beq{\begin{equation}}
\def\eeq{\end{equation}}

\def\bm{\begin{math}}
\def\me{\end{math}}

\begin{document}

\begin{center}
{\Large{\bf Phase Separation in Active Binary Mixtures With Chemical Reaction}} \\
\ \\
\ \\
by \\
Sayantan Mondal and Prasenjit Das\footnote{prasenjit.das@iisermohali.ac.in} \\
Department of Physical Sciences, Indian Institute of Science Education and Research -- Mohali, Knowledge City, Sector 81, SAS Nagar 140306, Punjab, India. \\
\end{center}

\begin{abstract}
\noindent We study motility-induced phase separation~(MIPS) in active AB binary mixtures undergoing the chemical reaction $A \rightleftharpoons B$. Starting from the evolution equations for the density fields $\rho_i(\vec r, t)$ describing MIPS, we phenomenologically incorporate the effects of the reaction through the reaction rate $\Gamma$ into the equations. The steady-state domain morphologies depend on $\Gamma$ and the relative activity of the species, $\Delta$. For a sufficiently large $\Gamma$ and $\Delta\ne 1$,  the more active component of the mixture forms a droplet morphology. We characterize the morphology of domains by calculating the equal-time correlation function $C(r, t)$ and the structure factor $S(k, t)$, exhibiting scaling violation. The average domain size, $L(t)$, follows a diffusive growth as $L(t)\sim t^{1/3}$ before reaching the steady state domain size, $L_{\rm ss}$. Additionally, $L_{\rm ss}$ shows the scaling relation $L_{\rm ss}\sim\Gamma^{-1/4}$, independent of $\Delta$.
\end{abstract}

\newpage
\section{Introduction}\label{sec1}
Active matter systems comprise large assemblies of individual units that can convert ambient energy into irreversible directed motion~\cite{MJSTJR13,S10,SYJ22}. Active matter encompasses a diverse array of systems with constituent sizes ranging from a few millimeters to several centimeters. Examples include bacterial colonies~\cite{DMA18}, autophoretic colloids~\cite{BDM17}, actin filaments~\cite{LCEA18}, bird flocks~\cite{AI24}, fish schools~\cite{MARYHW23}, and human crowds~\cite{AF23}. Unlike passive systems, active systems break the fluctuation-dissipation theorem and violate detailed balance at the microscopic level. Their inherently out-of-equilibrium nature gives rise to a diverse range of emergent phenomena, including active turbulence~\cite{RJJ22}, spatio-temporal pattern formation~\cite{BD17}, collective motion without attractive interactions, and active phase separation~\cite{JR19}.

A system of motile active particles can phase separate into dense and dilute phases~\cite{MJ15,JS21,OSTZ21}. This occurs because the particles slow down in regions of high accumulation. As particles gather and the local density increases, their mobility decreases further, creating a positive feedback loop. This mechanism ultimately leads to motility-induced phase separation (MIPS), where distinct dense and dilute regions form. In recent years, MIPS has been extensively studied in both single-component systems and binary mixtures, using theoretical and numerical approaches~\cite{STMAC2013,RAJRDM14,ANYCAJJ20,SSS21,SSS2}. Researchers have employed both microscopic and coarse-grained models to understand this phenomenon~\cite{CPDLG20,AKTP21,YJRB23,AJAYPJ23,DJMECJF21}. The first coarse-grained study of MIPS was due to Tailleur and Cates~\cite{JM08}. They demonstrated that under certain conditions, a system of run-and-tumble particles interacting through quorum sensing can separate into dense and dilute phases. Molecular simulations with various interaction potentials have also confirmed the occurrence of MIPS~\cite{GMA13,KS24}. Apart from theoretical works, numerous experiments across different systems have also validated the presence of MIPS. However, a detailed study of MIPS in active binary mixtures under the influence of chemical reactions remains an elusive area of research, despite extensive studies on its passive counterpart.

Before delving into the details, let's briefly summarize the phase separation kinetics in passive immiscible binary mixtures~\cite{JJ58,PW09} undergoing spinodal decomposition in the presence of chemical reactions. When a homogeneous symmetric AB binary mixture (comprising 50\%A and 50\%B) is quenched below its critical temperature $T_c$, it falls out of equilibrium. The subsequent evolution of the system is governed by the emergence of A-rich and B-rich domains. In the absence of reaction, the average domain size $L(t)$ increases with time following the Lifshitz-Slyozov (LS) law, characterized by $L(t)\sim t^{1/3}$~\cite{LS61}. The domain morphology is characterized by equal-time correlation function $C(r,t)$ and its Fourier transform $S(k,t)$, both of which exhibit dynamical scaling as $C(r, t)=g(r/L(t))$ and $S(k,t)=L(t)^df(kL(t))$. Here $g(x)$ and $f(p)$ are scaling functions, and $d$ is the spatial dimensionality~\cite{HH77,DP04}.

The first theoretical study of phase separation involving chemical reactions was due to Huberman~\cite{B76}. He investigated the formation of striations in a phase separating binary mixture simultaneously undergoing an autocatalytic chemical reaction. Glotzer \textit{et al.} performed Monte Carlo~(MC) simulations to investigate phase separation in AB binary mixtures undergoing the chemical reaction $A \rightleftharpoons B$~\cite{SDN94}. They showed that the inclusion of a chemical reaction leads to the formation of a unique steady-state pattern. The steady-state domain size $L_{\rm ss}$ scales with the reaction rate $\Gamma$ as $L_{\rm ss}\sim \Gamma^{-s}$, where $s=1/3$. This exponent matches the domain growth exponent $\alpha=1/3$ observed in systems without chemical reactions~\cite{LS61}. They further investigated the problem using a coarse-grained model and found that chemical reactions suppress the initial long-wavelength instability of spinodal decomposition, limiting domain growth to an intermediate size even in the late stages of phase separation.~\cite{SEM95}. Next, Puri and Frisch developed a model based on the master equation to examine the segregation dynamics in AB binary mixtures with chemically active components for two types of reactions: (i)~$A \rightleftharpoons B$  and (ii)~$A+B \rightleftharpoons 2B$~\cite{SH94,SH98}. The first molecular dynamics study of phase separation kinetics in a chemically reacting fluid mixture (for reaction type (i)) was due to Krishnan and Puri~\cite{RS15}. Unlike in solids, where domain growth is governed by diffusion, in fluids, the presence of a velocity field significantly influences the kinetics of domain growth. In their study, they selected reaction rates that lead to a steady-state domain structure forming within the viscous hydrodynamic regime, where the domain size grows as $L(t)\sim t$ and $L_{\rm ss}$ scales with $\Gamma$ as $L_{\rm ss}\sim \Gamma^{-1}$.

In all the previously discussed studies, the chemical reaction is driven by external agents, resulting in phase separation dynamics that are independent of reaction kinetics. Carati and Lefever were the first to study phase separation in binary mixtures with spinodal decomposition intrinsically coupled to chemical reactions~\cite{DR97}.  Following their pioneering work, numerous other research groups have explored the effect of this coupling on phase separation~\cite{TM22,TFEM23}.

The growing interest in active phase separation in the presence of chemical reactions is due to recent surge in research on proliferating active matter, which studies active particles with non-conserved dynamics like those found in biological systems~\cite{OSKJJBS23,D22,FC2024,ACF14}. J\"{u}licher and Weber reviewed a coarse-grained approach that captures the physics of chemically active emulsions, serving as a model for how condensates orchestrate chemical processes~\cite{CDFC19}. Zwicker \textit{et al.} studied active emulsions using a coarse-grained description of the droplet dynamics for two different chemical reactions~\cite{DAF15}. For first-order reactions, they observed that stable, monodisperse emulsions can form where Ostwald ripening is suppressed across a range of chemical reaction rates, even in the presence of thermal fluctuations. Conversely, in the case of autocatalytic droplets, spontaneous nucleation is significantly hindered, leading to typically unstable emulsions. However, these emulsions can be stabilized, and reliable nucleation of autocatalytic droplets can be achieved with the presence of chemically active cores. Cates \textit{et al.} discovered that birth and death processes coupled with MIPS can lead to arrested phase separation, relevant to pattern formation in concentrated biofilms and other biological systems~\cite{MDIJ11}. Li and Cates investigated non-equilibrium phase-separating systems with chemical reactions~\cite{YM21}. They found that the dynamics of the order parameter field closely resemble those of an equilibrium system with long-range attractive interactions. 

In this paper, we study MIPS in active AB binary mixtures undergoing the chemical reaction $\text{A} {\rightleftharpoons} \text{B}$. Motility facilitates phase separation into A-rich and B-rich domains, while the reaction counteracts this by homogenizing the compositions, resulting in complex domain kinetics. Our primary goal is to examine the domain morphologies that arise from these complex kinetics, their dependence on the relative activity of the species and reaction rate, and the presence of dynamical scaling. Given this, the paper is organized as follows: Section~\ref{sec2} presents the details of the modeling and analysis. The simulation details and results are discussed in Sec.~\ref{sec3}. Finally, the outcomes are summarized in Sec.~\ref{sec4}.

\section{Details of Modeling and Analysis}\label{sec2}
We begin with a coarse-grained model of active run-and-tumble particles in the presence of quorum sensing interactions in $d=2$. In this model, the fluctuating density field of the $i$th species $\rho_i(\vec r) = \sum_\alpha\delta(\vec r - \vec r_{i,\alpha})$ represents a hydrodynamic mode. For a system with  $N$-species, it has been shown that the stochastic dynamics of the density fields $\rho_i(\vec r, t)$ can be described by $N$ coupled It\"o-Langevin equations of the form~\cite{CuraT,KGS23,SMPD24}
\begin{eqnarray}
\label{eqn1}
\frac{1}{2\beta}\frac{\partial\rho_{i}(\vec r,t)}{\partial t}=\vec\nabla\cdot\left[v_{i}^2\vec\nabla\rho_{i}(\vec r,t)+v_{i}\rho_{i}(\vec r,t) \vec\nabla v_{i} + \sqrt{\rho_i/\beta}v_i\vec\Lambda_i(\vec r, t) \right]- \zeta\nabla^4\rho_{i}(r,t).
\end{eqnarray}
Here $v_i$ is the speed of $i$th species and $\beta$ is the tumbling rate. The Gaussian thermal noise $\vec\Lambda_i(\vec r, t)$ satisfies $\langle\Lambda_{i,\mu}(\vec r, t)\rangle=0$ and $\langle\Lambda_{i,\mu}(\vec r_1, t_1)\Lambda_{j,\nu}(\vec r_2, t_2)\rangle=\delta_{ij}\delta_{\mu\nu}\delta(\vec r_1 - \vec r_2)\delta(t_1-t_2)$, where $\mu$ and $\nu$ refer to Cartesian coordinates. We have phenomenologically added the term $\zeta\nabla^4\rho_i$ to stabilize domain interfaces between different species, similar to approaches in passive systems~\cite{PW09,DP04}. Here, $\zeta$ stands for the strength of the surface tension, and we set $\zeta>0$ to induce phase separation. This model is motivated by its relevance to bacterial colonies where quorum-sensing behavior is well established~\cite{ANYCAJJ20,CuraT}.

We use the following form of speeds $v_i$ of the species
\begin{eqnarray}
\label{eqn2}
v_i(\{\rho_{n \neq i}\}) = v_{i,0} \exp\left( \lambda \prod_{n \neq i} \rho_n\right),
\end{eqnarray}
where $\lambda$ is the interaction strength among the species and $v_{i,0}$ represents the amplitude of the speed $v_i$. These speeds represent a specific form of quorum-sensing interaction, where the speed of one species is regulated by the density of the other species. Tailleur and Cates have shown that the above form of speeds maps the system to an equilibrium system, whose dynamics can be obtained from a free energy functional~\cite{ANYCAJJ20,CuraT}. For an active AB binary mixture, eq.~(\ref{eqn2}) yields
\begin{eqnarray}
\label{eqn3}
v_{\rm A}(\rho_{\rm B})=v_{\rm A,0}\exp(\lambda\rho_{\rm B}) ~~~\text{and}~~~v_{\rm B}(\rho_{\rm A})=v_{\rm B,0}\exp(\lambda\rho_{\rm A}).
\end{eqnarray}
The free energy density of the system is given by
\begin{eqnarray}
\label{eqn4}
f(\rho_{\rm A},\rho_{\rm B}) = \rho_{\rm A} (\ln \rho_{\rm A} - 1) + \rho_{\rm B} (\ln \rho_{\rm B} - 1) + \lambda\rho_{\rm A}\rho_{\rm B}.
\end{eqnarray}
The first two terms on RHS correspond to the free energy density of an ideal gas, and the last term represents the excess free energy density due to mutual interaction~\cite{CuraT,SMPD24}. The concavity condition on $f(\rho_{\rm A},\rho_{\rm B})$ provides the instability condition as
\begin{eqnarray}
\label{eqn5}
\lambda^2>\frac{1}{\rho_{\rm A}\rho_{\rm B}}.
\end{eqnarray}
The sign of the parameter $\lambda$ determines the nature of interaction between the two species: when $\lambda<0$, the species mutually inhibit each other, resulting in colocalization, while $\lambda>0$ indicates mutual activation, leading to phase separation~\cite{ANYCAJJ20,CuraT}. 

Next, we consider a reversible reaction process where species A can convert to species B and vice versa, with the equal rate of conversion $\Gamma^\prime$ in both directions, represented as
\begin{eqnarray}
\label{eqn6}
\text{A} \overset{\Gamma^\prime}{\underset{\Gamma^\prime}{\rightleftharpoons}} \text{B}.
\end{eqnarray}
Using the framework of chemical rate equations, we have modified the evolution equations~(\ref{eqn1}) to incorporate this reaction as follows:
\begin{eqnarray}
\label{eqn7}
\frac{1}{2\beta}\frac{\partial\rho_{\rm A,B}}{\partial t} = \vec{\nabla} \cdot \left[ v_{\rm A,B}^2 \vec{\nabla} \rho_{\rm A,B} + v_{\rm A,B} \rho_{\rm A,B} \vec{\nabla} v_{\rm A,B} + \sqrt{\rho_{\rm A,B}/\beta}v_{\rm A,B}\vec\Lambda_{\rm A,B} \right] - \zeta\nabla^4 \rho_{\rm A,B} \mp \Gamma(\rho_{\rm A} - \rho_{\rm B}),
\end{eqnarray}
where $\Gamma^\prime=2\beta\Gamma$. Here, we emphasize that we consider the phase separation dynamics as decoupled from the reaction kinetics. This scenario is achievable when the reaction is controlled by external agents. Additionally, we do not include any extra noise term related to the reactive part of the dynamics, as the chemical reaction is treated in a mean-field approximation.

To determine whether the above chemical reaction affects the instability condition given by eq.~(\ref{eqn5}), we perform a linear stability analysis of the governing equations~(\ref{eqn7}). To proceed, we neglect the noise term, as we are interested in the mean-field limit. Let $\rho_i = \rho_{i,0} + \delta \rho_i$, where $\delta \rho_i$ represents a small perturbation around the homogeneous initial density $\rho_{i,0}$ of $i$th species ($i\in A,B$). The expansion of eqs.~(\ref{eqn7}) to a linear order in $\delta \rho_i$ provides
\begin{eqnarray}
\label{eqn12a}
\frac{1}{2\beta}\Dot{\delta\rho_{\rm A}} &=& \left[v_{\rm A,0}^2\nabla^2\delta\rho_{\rm A} + \lambda v^2_{\rm A,0}\rho_{\rm A,0} \nabla^2 \delta\rho_{\rm B} \right] - \Gamma(\delta\rho_{\rm A} - \delta\rho_{\rm B}) - \zeta q^4 \delta\rho_{\rm A} + O(\delta\rho_{\rm A}^2, \delta\rho_{\rm B}^2,\delta\rho_{\rm A}\delta\rho_{\rm B}), \\
\label{eqn12b}
\frac{1}{2\beta}\Dot{\delta\rho_{\rm B}} &=& \left[v_{\rm B,0}^2\nabla^2\delta\rho_{\rm B} + \lambda v^2_{\rm B,0}\rho_{\rm B,0} \nabla^2 \delta\rho_{\rm A} \right] + \Gamma(\delta\rho_{\rm A} - \delta\rho_{\rm B}) - \zeta q^4 \delta\rho_{\rm B} + O(\delta\rho_{\rm A}^2, \delta\rho_{\rm B}^2,\delta\rho_{\rm A}\delta\rho_{\rm B}).
\end{eqnarray}
Introducing the Fourier transformation of the density perturbation as
\begin{eqnarray}
\label{eqn13}
\delta\rho_i^{\vec q}= \int d\vec r \, e^{-i \vec q \cdot \vec r}\delta\rho_i,
\end{eqnarray}
we write eqs.~(\ref{eqn12a}) and (\ref{eqn12b}) in Fourier space:
\begin{eqnarray}
\label{eqn14}
\delta{\dot \rho}^{\vec{q}} = - 2\beta M_q \delta\rho^{\vec q}.
\end{eqnarray}
Here $\delta\rho^{\vec q}\equiv (\delta\rho_A^{\vec q}, \delta\rho_B^{\vec q})$ is a column vector and the dynamical matrix $M_q$ is given by:
\begin{eqnarray}
\label{eqn15}
M_q = \begin{bmatrix}
\rule{0pt}{2.5ex} q^4\zeta + q^2 v^2_{\rm A,0} + \Gamma &  q^2\lambda v^2_{\rm A,0}\rho_{\rm A,0} -\Gamma \\ \rule{0pt}{2.5ex}
q^2 \lambda v^2_{\rm B,0}\rho_{\rm B,0}- \Gamma &  q^4\zeta + q^2v^2_{\rm B,0} +\Gamma \rule{0pt}{2.5ex}
\end{bmatrix}
\end{eqnarray}
An initial homogeneous state is linearly stable if $\text{Tr}(M_q)>0$ and $\text{det}(M_q)>0$. If either condition is violated, a linear perturbation will grow exponentially, leading to phase separation. Since $\text{Tr}(M_q)$ is always positive, the instability criterion is $\text{det}(M_q)<0$. Carrying out the calculations, one obtains:
\begin{eqnarray}
\label{eqn16}
\text{det}(M_q)&=&q^4 \left\{v^2_{\rm A,0}v^2_{\rm B,0}\left(1-\lambda^2\rho_{\rm A,0}\rho_{\rm B,0}\right) + 2\zeta\Gamma \right\} + q^2\Gamma\left\{v^2_{\rm A,0}\left(1+\lambda\rho_{\rm A,0}\right)+v^2_{\rm B,0}\left(1+\lambda\rho_{\rm B,0}\right) \right\} \nonumber \ \\ &+& \zeta q^6(v^2_{\rm A,0} + v^2_{\rm B,0}) + \zeta^2 q^8.
\end{eqnarray}
As all the terms on the right-hand side of eq.~(\ref{eqn16}) are always positive except for the first term, the first term must be negative to satisfy the instability criterion, leading to the following condition:
\begin{eqnarray}
\label{eqn16a}
v^2_{\rm A,0}v^2_{\rm B,0}\left(1-\lambda^2\rho_{\rm A,0}\rho_{\rm B,0}\right) + 2\zeta\Gamma < 0.
\end{eqnarray}
Since $2\zeta\Gamma>0$, eq.~(\ref{eqn16a}) provides the necessary condition for instability:
 \begin{eqnarray}
\label{eqn17}
\lambda^2>\frac{1}{\rho_{\rm A,0}\rho_{\rm B,0}},
\end{eqnarray}
which is same as the condition of MIPS in absence of reaction, given by eq.~(\ref{eqn5}).

To determine the critical reaction rate $\Gamma_c$, above which the system does not exhibit phase separation despite satisfying the condition in eq.~(\ref{eqn17}), we compute the eigenvalues of the stability matrix $M_q$, given by
\begin{eqnarray}
\omega_\pm (q^2) = \frac{\text{Trace} (M_q) \pm \sqrt{(\text{Trace}(M_q))^2 - 4 \text{Det}(M_q)}}{2}.
\end{eqnarray}
In this case, the smallest eigenvalue, $\omega_-(q^2)$, represents the dominant growth rate. The fastest growing mode, or the least stable mode, $q_m$, can be obtained from the condition
\begin{eqnarray}
\left.\frac{d\omega_{-} (q^2)}{dq^2}\right|_{q=q_m} = 0.
\end{eqnarray}
Finally, $\Gamma_c$ is determined from
\begin{eqnarray}
\left. \text{Det}(M_q)\right|_{\Gamma=\Gamma_c,q=q_m} = 0.
\end{eqnarray} 
The eq.~(\ref{eqn16a}) can only provide a rough estimate of $\Gamma_c$, when the condition of MIPS is satisfied.

Next, we non-dimensionalize eq.~(\ref{eqn7}) together with eqs.~(\ref{eqn3}) and (\ref{eqn5}) by rescaling the variables as follows:
$$v_i = v_0\Tilde{v_i},~~ \vec r = \sqrt{\zeta/(v_0)^2}\Tilde{\vec r},~~ t = \frac{\zeta}{2\beta (v_0)^4}\Tilde{t},~~\Gamma^\prime=\frac{2\beta (v_0)^4}{\zeta}\Tilde{\Gamma},~~\rho_i=\frac{\Tilde{\rho_i}}{|\lambda|},~~ \vec\Theta_{i}=\frac{v_0^3}{|\lambda|\sqrt{\zeta}}\tilde{\vec\Theta}_{i}.$$
Here, all the variables with a tilde are dimensionless quantities, $v_0$ is a scale of speed, and $i\in(A,B)$. In terms of rescaled variables (dropping tildes), eq.~(\ref{eqn7}) reduces to
\begin{eqnarray}
\label{eqn8}
\frac{\partial\rho_{\rm A,B}}{\partial t}=\vec\nabla\cdot\left[v_{\rm A,B}^2\vec\nabla\rho_{\rm A,B}+v_{\rm A,B}\rho_{\rm A,B}(\vec r,t) \vec\nabla v_{\rm A,B}+\vec{\Theta}_{\rm A,B}\right]- \nabla^4\rho_{\rm A,B}\mp \Gamma(\rho_{\rm A} - \rho_{\rm B}),
\end{eqnarray}
Further, we set $v_A^0=v_0$ and $v_B^0=v_0\Delta$, where $\Delta$ represents the relative activity of the two species. Consequently, eq.~(\ref{eqn3}) changes to
\begin{eqnarray}
\label{eqn9}
v_{A}(\rho_B)=\exp(\pm\rho_B) ~~~\text{and}~~~v_{B}(\rho_A)=\Delta\exp(\pm\rho_A).
\end{eqnarray}
The + and - signs correspond to phase separation and colocalization of the species, respectively. The Gaussian white noise $\vec{\Theta_{i}}(\vec{r},t)$ obeys the following relations in $d$-dimension:
\begin{eqnarray}
\label{eqn10}
\langle\vec\Theta_i(\vec{r},t)\rangle=0~\text{and}~
\langle\Theta_{i,\mu}(\vec{r}_1, t_1)\Theta_{j,\nu}(\vec{r}_2, t_2)\rangle=\epsilon \rho_i v_i^2\delta_{ij} \delta_{\mu\nu} \delta(\vec{r}_1 - \vec{r}_2) \delta(t_1 - t_2)~\text{with}~\epsilon = \frac{d|\lambda|v_0^d}{\zeta^{\frac{d}{2}}},
\end{eqnarray} 
where $i$ and $j$ stand for species. Finally, the condition of instability given by eq.~(\ref{eqn16a}) reduces to
\begin{eqnarray}
\label{eqn31}
\Delta^2\left(1-\rho_{\rm A,0}\rho_{\rm B,0}\right) + 2\Gamma < 0,
\end{eqnarray}
and the necessary condition for instability given by eq.~(\ref{eqn5}) or eq.~(\ref{eqn17}) becomes
\begin{eqnarray}
\label{eqn11}
\rho_A\rho_B>1.
\end{eqnarray}

\begin{figure}
\centering
\includegraphics*[width=0.50\textwidth]{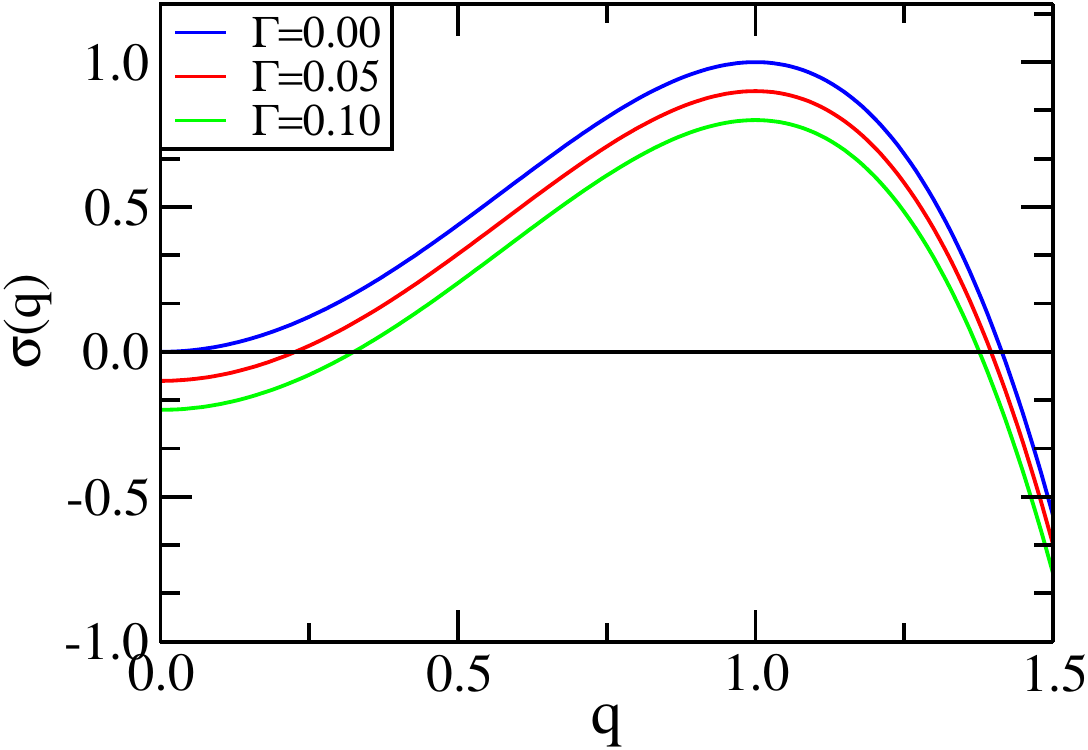}
\caption{\label{fig0} Plot of $\sigma(q)$ vs. $q$ for $q_u=\sqrt{2}$ and different values of $\Gamma$, as mentioned. For nonzero $\Gamma$, $\sigma(q)$ becomes negative at $q=0$.}
\end{figure}

Next, we perform the linear stability analysis of eq.~(\ref{eqn8}), using $\delta\phi = \delta\rho_A - \delta\rho_B$. The dynamical equation for $\delta\phi$ is given by
\begin{eqnarray}
\label{eqn19}
\delta\dot{\phi} = v_0^2(1-\rho_0)\nabla^2\delta\phi - \nabla^4\delta\phi -2\Gamma\delta\phi.
\end{eqnarray}
Here, we consider same initial density $\rho_{0}$ for both the species and assume $v^2_{\rm A,0}=v^2_{\rm B,0}=v^2_{0}$. The latter is particularly true for $\Delta=1$ case. The Fourier transform of eq.~(\ref{eqn19}) provides 
\begin{eqnarray}
\label{eqn20}
\dot{\delta\phi_q}=\left\{q^2\left(q^2_u-q^2\right)-2\Gamma  \right\}\delta\phi_q,
\end{eqnarray}
where $q_u=\sqrt{(\rho_0-1)v^2_{0}}$. Eq.~(\ref{eqn20}) shows that the $q^4$ term from surface tension stabilizes the domain morphology without affecting the stability criteria, as mentioned earlier. The solution of eq.~(\ref{eqn20}) is given by
\begin{eqnarray}
\label{eqn21}
{\delta\phi_q(t)}=\delta\phi_q(0)\exp\{\sigma(q)t\},
\end{eqnarray}
where $\sigma(q)=q^2\left(q^2_u-q^2\right)-2\Gamma$ is the growth factor. For $\Gamma=0$, $\sigma(q)$ has a cut-off at $q=0$ and $q=q_u$. Fluctuations with $q > q_u$ decay, while those with $0 < q < q_u$ grow over time with the maximum growth rate occurring for $q_p=q_u/\sqrt{2}$. Any nonzero $\Gamma$ shifts the cut-offs at both small and large $q$, allowing growth only for intermediate wave number fluctuations, as shown in Fig.~{\ref{fig0}}. Thus, the reaction suppresses concentration fluctuations at small wave numbers, resulting in the freezing of domain patterns.

\section {Numerical Simulations And Results}\label{sec3}
We numerically solve eq.~(\ref{eqn8}) using the Euler discretization method in $d=2$ for 50\%-50\% binary mixtures of species A and B, considering different relative activities $\Delta$ and reaction rates $\Gamma$. The speed of $i$th species ($i\in A,B$) $v_i$ is obtained by using eq.~(\ref{eqn9}) with + sign. Initially, the densities of the species, $\rho_i$, are set as small fluctuations around the mean density $\rho_0$, such that $\rho_i = \rho_0 \pm 0.01$, where $\rho_0 = 1.065$. This corresponds to a homogeneous binary mixture at $t=0$. Clearly, our choice of densities satisfies the necessary instability condition as specified in eq.~(\ref{eqn11}). We set our system size to $L_x \times L_y = 512 \times 512$ for all cases. To ensure numerical stability, we use a spatial discretization with a mesh size of $\Delta x = 0.5$ and a time step of $\Delta t = 0.001$. We define the order parameter as $\psi(\vec r,t)=[\rho_{A}(\vec r,t) - \rho_{B}(\vec r,t)]/[\rho_{A}(\vec r,t) + \rho_{B}(\vec r,t)]$, where $\rho_i(\vec r,t)$ is the local concentration of $i$th species at position $\Vec{r}$ and time $t$. Consequently, regions with $\psi > 0$ correspond to A-rich domains, while regions with $\psi < 0$ correspond to B-rich domains. We impose periodic boundary conditions in all directions, and all statistical quantities are averaged over ten independent runs. Furthermore, we use $\epsilon = 0.02$ in all simulations, as changing $\epsilon$ does not impact the statistical results.

\begin{figure}
\centering
\includegraphics*[width=0.56\textwidth]{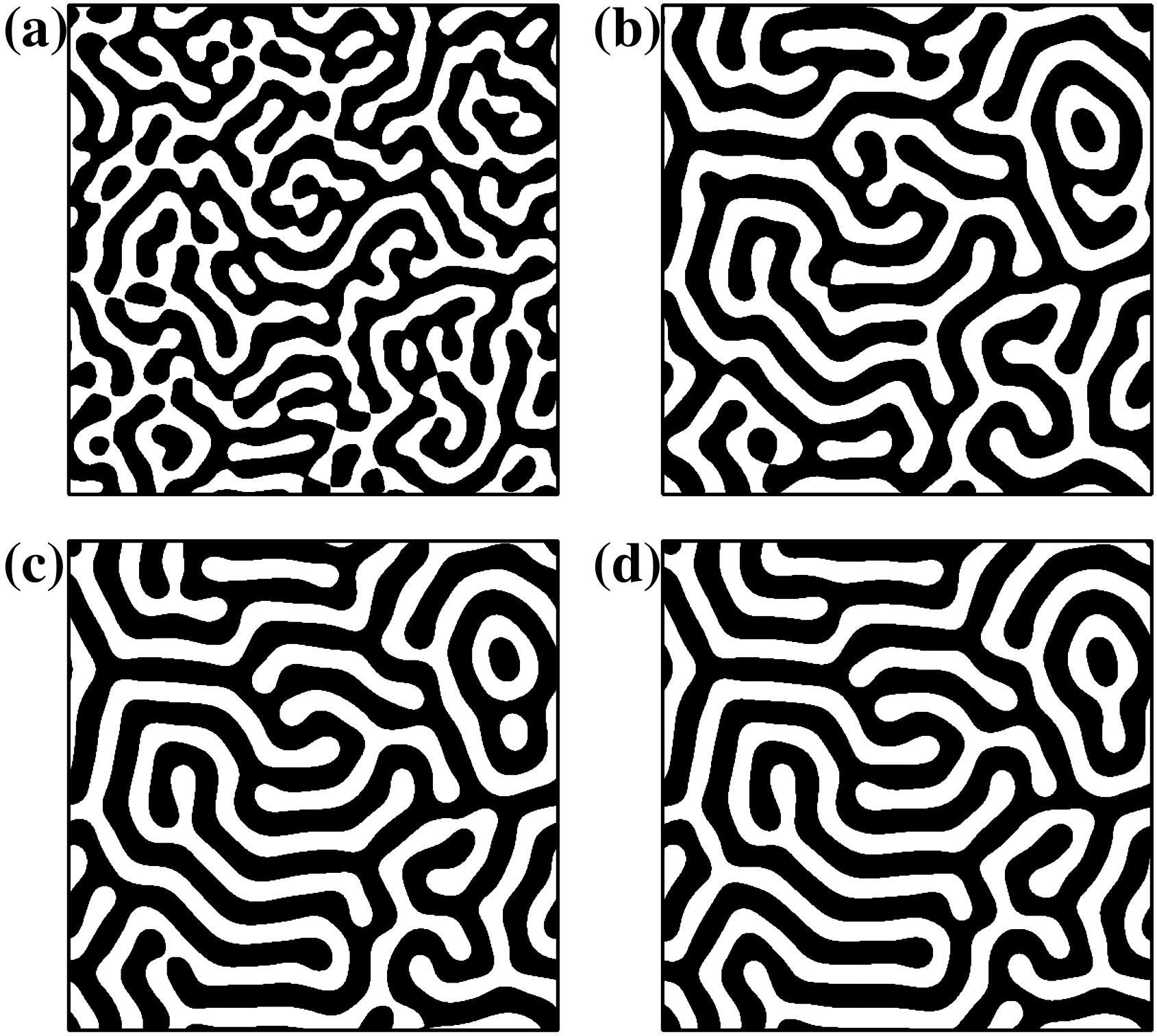}
\caption{\label{fig1} Evolution snapshots of an active binary mixture (50\%A - 50\%B) undergoing the chemical  reaction $\text{A} \overset{\Gamma}{\underset{\Gamma}{\rightleftharpoons}} \text{B}$ are shown at different times: (a) $t = 500$, (b) $t = 2000$, (c) $t = 5000$, and (d) $t = 10000$. The reaction rate is $\Gamma = 0.003$, and the relative activity is $\Delta = 1$. Regions with $\psi < 0$ are depicted in black, while regions with $\psi > 0$ remain unmarked.}
\end{figure}

Figure~\ref{fig1} shows the evolution snapshots of the order parameter field for $\Delta=1$ and $\Gamma=0.003$ at different times, as mentioned. At early times, we observe a characteristic bicontinuous domain morphology, resembling the patterns observed in MIPS without chemical reactions~\cite{SMPD24}, as shown in Fig.~\ref{fig1}(a). This indicates that the phase separation is primarily governed by bulk diffusion, with the chemical reaction playing a minimal role in influencing phase separation at these early stages. However, as time progresses, the ongoing chemical reaction begins to influence the process of phase separation, leading to the formation of a complex domain structure at intermediate stages, as shown in Fig.\ref{fig1}(b). With further progression, a transition to a labyrinthine pattern is observed, as presented in Fig.\ref{fig1}(c). Additionally, at the late stage of evolution, the domain structure appears to freeze, as theoretically predicted and shown in Figs.~\ref{fig1}(c) and \ref{fig1}(d). Here, we emphasize that the steady-state labyrinthine pattern persists for other values of $\Gamma$ as well, with $\Delta$ fixed at 1.

\begin{figure}
\centering
\includegraphics*[width=0.56\textwidth]{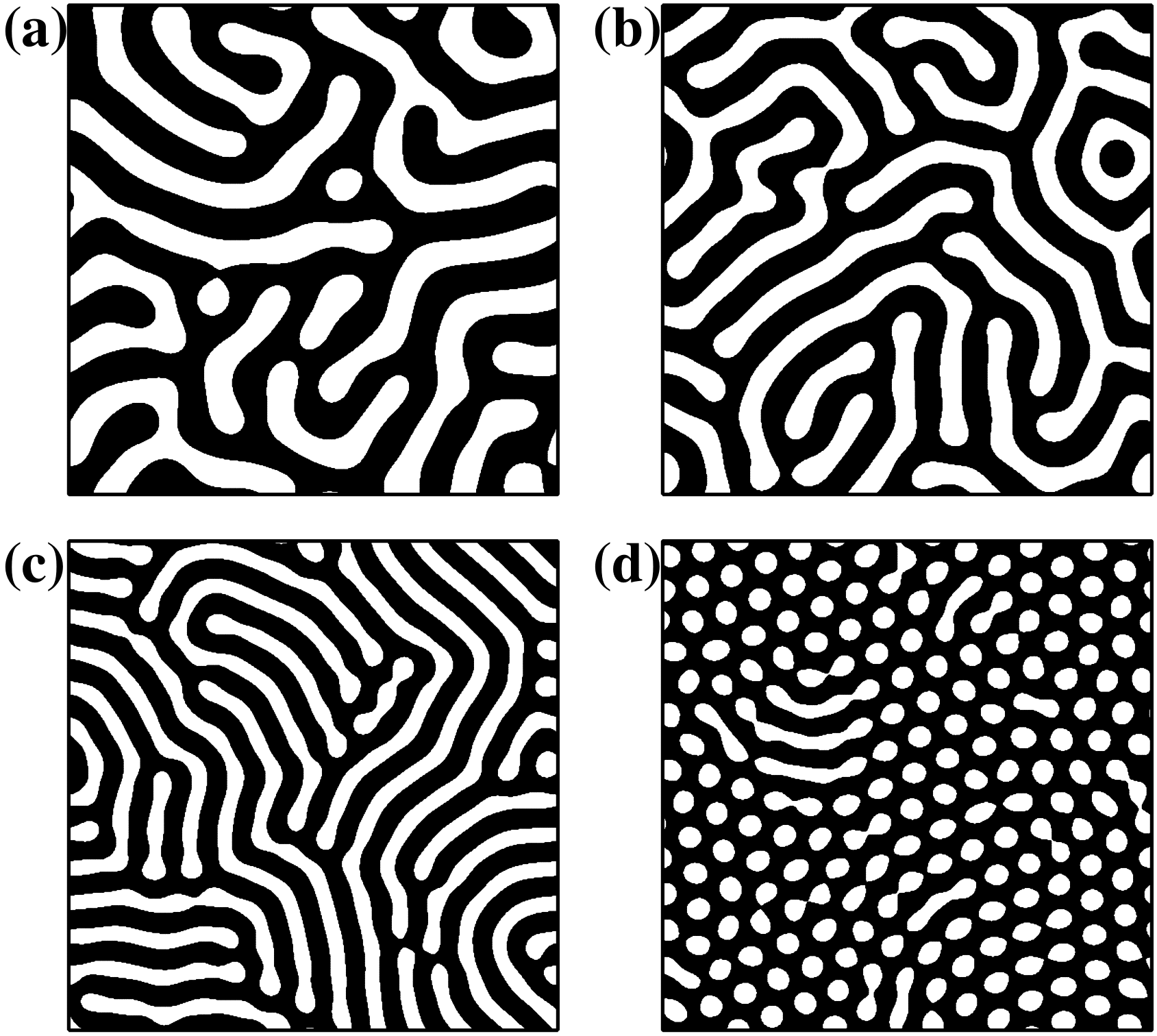}
\caption{\label{fig4} Evolution snapshots of an active binary mixture undergoing the reaction $\text{A} \overset{\Gamma}{\underset{\Gamma}{\rightleftharpoons}} \text{B}$ are shown at $t=8000$ for $\Delta=0.7$ and different reaction rates: (a) $\Gamma=0.0008$, (b) $\Gamma=0.002$, (c) $\Gamma=0.007$, and (d) $\Gamma=0.01$. Other details remain the same as in Fig.~\ref{fig1}.}
\end{figure}

Next, we investigate the dependence of steady-state domain morphology on the reaction rate $\Gamma$ under the condition $\Delta\ne 1$. Figure~\ref{fig4} shows the evolution snapshots of the systems at $t=8000$ for $\Delta=0.7$ and various values of $\Gamma$, as specified. For a small reaction rate, $\Gamma=0.0008$, the domain morphology exhibits a bicontinuous structure, as shown in Fig.~\ref{fig4}(a), similar to the patterns observed in MIPS without chemical reactions~\cite{SMPD24}. At this low $\Gamma$, the system has not yet reached a steady state by $t = 8000$. As $\Gamma$ increases, the chemical reactions begin to influence the domain morphology, resulting in a transition to a steady-state labyrinthine pattern, as shown in Figs.~\ref{fig4}(b) and \ref{fig4}(c). When $\Gamma$ is further increased to \(0.01\), the more active component of the mixture (A-component) forms a droplet-like morphology. Here, we emphasize that even at such high values of $\Gamma$, the system still exhibits phase separation. A pertinent question now arises: does the B-component exhibit droplet-like morphology at a large $\Gamma$ when $\Delta>1$?

\begin{figure}
\centering
\includegraphics*[width=0.56\textwidth]{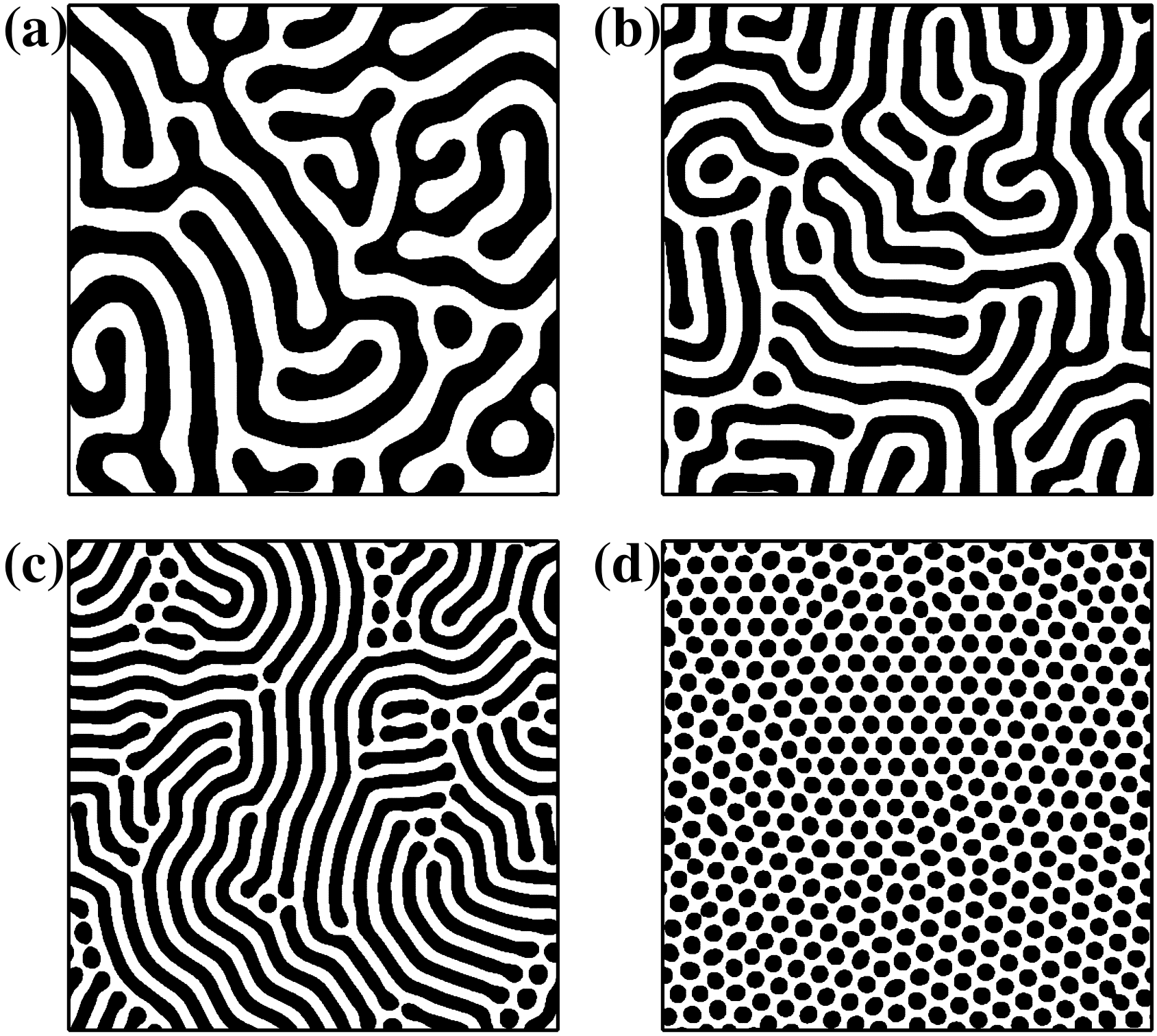}
\caption{\label{fig5} Figure analogous to Fig.~\ref{fig4} at $t=8000$ for $\Delta=2.0$ and various reaction rates: (a) $\Gamma=0.002$, (b) $\Gamma=0.007$, (c) $\Gamma=0.03$, and (d) $\Gamma=0.05$. Other details are the same as in Fig.~\ref{fig1}.}
\end{figure}

Figure~\ref{fig5} shows the evolution snapshots of the system at $t=8000$ for $\Delta=2.0$ and various values of $\Gamma$, as mentioned. Similar to $\Delta=0.7$, we see bicontinuous and labyrinthine pattern for small and intermediate range of $\Gamma$, as shown in Figs.~\ref{fig5}(a)-(c). However, at a larger value of $\Gamma=0.05$, the B-component of the mixture shows a droplet morphology. Thus, we can conclude that for large $\Gamma$, the component of the mixture with higher activity (when $\Delta\ne 1$) exhibits a droplet-like morphology.

\begin{figure}
\centering
\includegraphics*[width=0.85\textwidth]{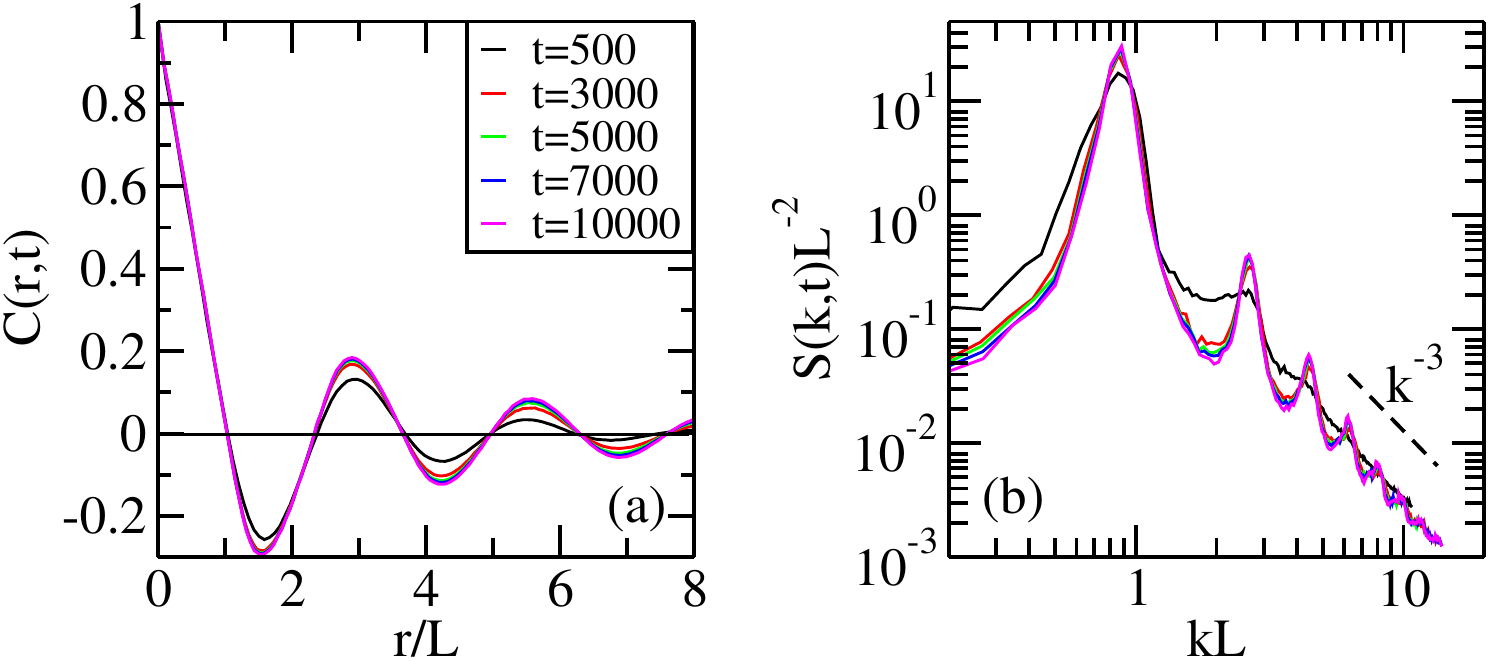}
\caption{\label{fig2} Scaling plots of correlation functions and structure factors corresponding to the evolution shown in Fig.~\ref{fig1}. (a) Plot of $C(r, t)$ vs. $r/L(t)$ at the indicated times. The length scale $L(t)$ is defined as the first zero-crossing of $C(r, t)$. (b) Log-log plot of $S(k, t)L^{-2}$ vs. $kL(t)$ at various times. The symbols have the same meaning as in (a). The dashed black line, labeled $k^{-3}$, represents \textit{Porod's law} for $d=2$.}
\end{figure}

Next, we characterize the morphological transition of the domains observed in Fig.~\ref{fig1} by computing the equal-time correlation function $C(\vec r, t)$ and structure factor $S(\vec k, t)$. For the order parameter $\psi(\vec r, t)$, the equal-time correlation function is defined as 
\begin{eqnarray}
\label{apeqn12}
C(r, t)=\bigl\langle\psi(\vec{R}+\vec{r}, t)\psi(\vec{R}, t)\bigr\rangle-\bigl\langle\psi(\vec{R}+\vec{r}, t) \bigr \rangle\bigl\langle\psi(\vec{R}, t)\bigr\rangle.
\end{eqnarray}
Here, $\langle\cdots\rangle$ denotes an average over reference positions $\vec{R}$ and spherical averaging over different directions~\cite{PW09,HH77,DP04}. The structure factor $S(\vec k, t)$ is defined as the Fourier transform of $C(\vec r, t)$ at wave vector $\vec k$ as 
\begin{eqnarray}
\label{apeqn13}
S(\vec k, t)  = \int d\Vec{r}  e^{i\Vec{k}.\Vec{r}} C(\Vec{r},t).
\end{eqnarray}
In Fig.~\ref{fig2}(a), we have shown a plot of $C(r, t)$ vs. $r/L(t)$ at various times, as mentioned. Here, $L(t)$ is the average domain size. It is defined as the distance at which the correlation function $C(r, t)$ first decays to zero from its maximum value at $r=0$. The numerical data at different times clearly collapse onto each other at small distances but deviate at larger distances. The oscillatory behavior of $C(r, t)$ at large distances becomes more pronounced over time before eventually settling into a steady state. This phenomenon arises from the formation of locally periodic domains in the labyrinthine structure. Therefore, we conclude that there is a violation of dynamical scaling, indicating that the domain structures at different times are not scale-invariant. In Fig.~\ref{fig2}(b), we have plotted $S(k, t)L^{-2}$ vs. $kL(t)$ at different times on a log-log scale. Similar to $C(r, t)$, $S(k, t)$ also exhibits dynamical scaling violation. At intermediate to later times, $S(k, t)$ shows an oscillatory decay, resulting from the presence of locally periodic domains. This is in stark contrast to its early-time behavior (e.g., $S(k, t)$ at $t=500$, represented by the solid black line in Fig.~\ref{fig2}(b)), where it decays smoothly without oscillations. However, the tail of $S(k, t)$ decays as $k^{-3}$ in the limit $k \rightarrow \infty$, consistent with the well-known \textit{Porod's Law}, which states that $S(k, t) \sim k^{-(d+1)}$ in $d=2$ at all times~\cite{GP82,A94}. This behavior confirms scattering from the sharp interfaces between A-rich and B-rich domains. Additionally, these results remain valid for other values of $\Gamma$ and $\Delta$.

\begin{figure}
\centering
\includegraphics*[width=0.55\textwidth]{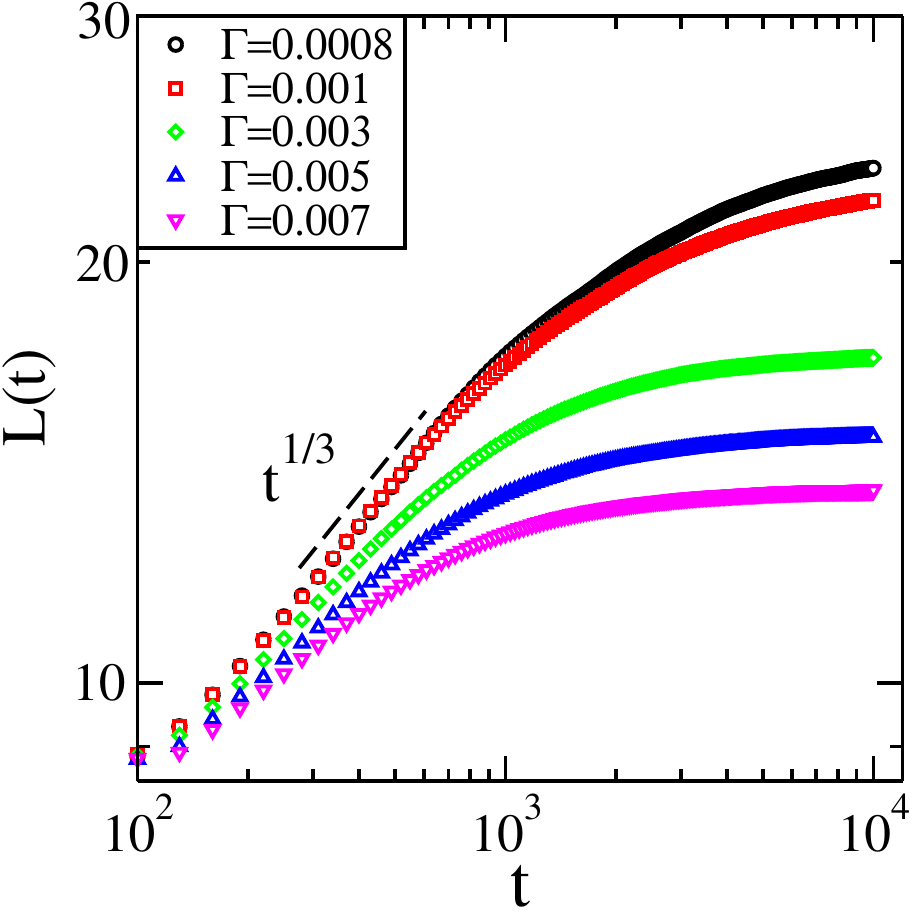}
\caption{\label{fig3} Plot of average domain size $L(t)$ vs. $t$ on a log-log scale for $\Delta=1$ and various values of $\Gamma$, as indicated. The dashed black line labeled $t^{1/3}$ represents the LS law: $L(t)\sim t^{1/3}$.}
\end{figure}

Figure~\ref{fig3} shows a log-log plot of $L(t)$ vs. $t$ for $\Delta=1$ and various values of $\Gamma$. At early times, $L(t)$ exhibits power-law growth as $L(t) \sim t^{1/3}$, following the LS law~\cite{DP04,LS61}. This confirms that, at early stages, phase separation is primarily driven by bulk diffusion, similar to the behavior observed in systems without chemical reactions~\cite{SMPD24}. However, at the late stage of evolution, when chemical reactions play a significant role, $L(t)$ reaches a steady-state value, $L_{\rm ss}$. Further, the time $t_{\rm s}$ at which $L(t)$ attains $L_{\rm ss}$, as well as the value of $L_{\rm ss}$ itself, depends on $\Gamma$ for a fixed $\Delta$. A higher value of $\Gamma$ results in a smaller $t_{\rm s}$.

\begin{figure}
\centering
\includegraphics*[width=0.55\textwidth]{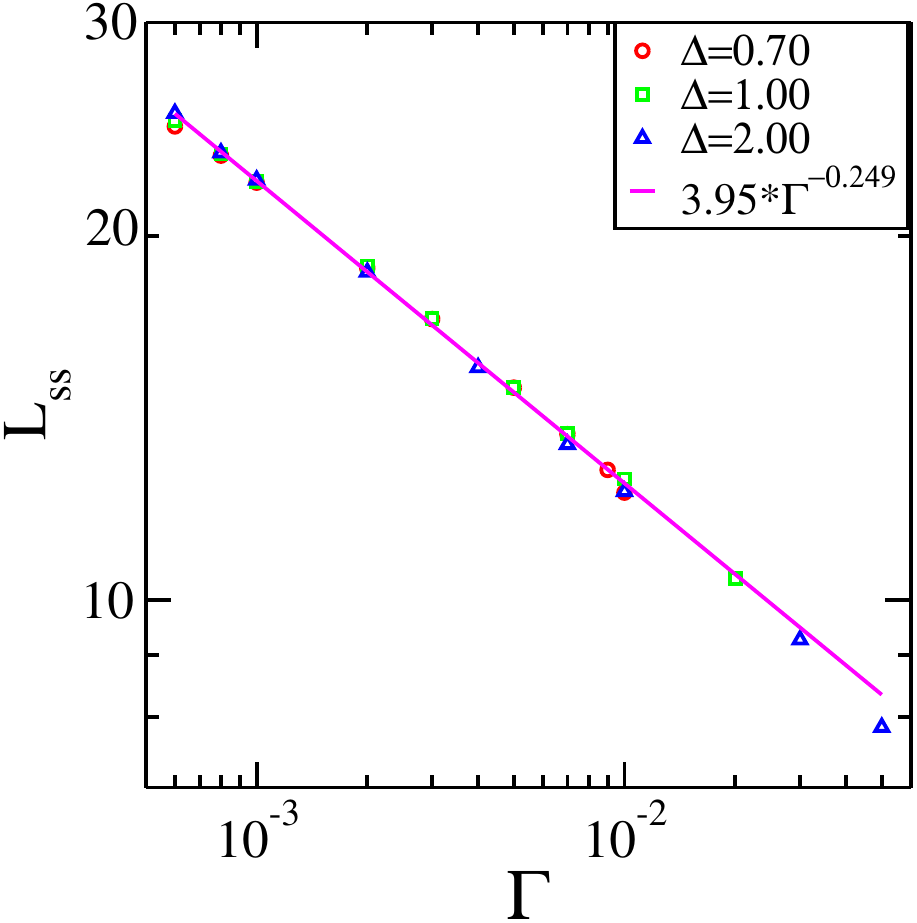}
\caption{\label{fig6} Plot of $L_{\rm ss}$ vs. $\Gamma$ on a log-log scale for different values of $\Delta$, as indicated. The solid magenta line, $L_{\rm ss}=3.95\Gamma^{-0.249}$, represents the best fit to our numerical data.}
\end{figure}

Finally, we examine the dependence of $L_{\rm ss}$ on $\Gamma$. Figure~\ref{fig6} shows a log-log plot of $L_{\rm ss}$ vs. $\Gamma$ for different values of $\Delta$, as mentioned. The best fit of our numerical data reveals that $L_{\rm ss}\sim\Gamma^{-1/4}$, irrespective of $\Delta$. This result stands in clear contrast to passive binary mixtures, where $L_{\rm ss}\sim\Gamma^{-\alpha}$, with $L(t)$ approaching $L_{\rm ss}$ following the growth law: $L(t)\sim t^{\alpha}$~\cite{SDN94,SEM95,RS15}. Here, although $L(t)$ reaches $L_{\rm ss}$ following the LS law: $L(t)\sim t^{1/3}$, the steady-state scaling exhibits $L_{\rm ss}\sim\Gamma^{-1/4}$.

\section{Summary and Discussion}\label{sec4}
In summary, we have studied motility-induced phase separation~(MIPS) in active AB binary mixtures in presence of the chemical reaction $\text{A} \overset{\Gamma}{\underset{\Gamma}{\rightleftharpoons}} \text{B}$ in $d=2$. Starting with the coarse-grained evolution equations for the density fields $\rho_i(\vec r,t)$, we phenomenologically incorporated the effects of the reaction into these equations. Our linear analysis of evolution equations revealed that the necessary condition for instability is unchanged by the presence of reactions. For an interaction parameter $\lambda > 0$ that satisfies the instability condition, phase separation is prevented if the reaction rate exceeds a critical value, $\Gamma_c$. Moreover, we found that the reaction suppresses concentration fluctuations at small wave numbers, resulting in the freezing of domains at late times.

At $t=0$, we begin with a homogeneous 50\%A-50\%B mixture with the interaction parameter $\lambda > 0$, satisfying the necessary instability condition. As time increases, the system undergoes phase separation into A-rich and B-rich regions through diffusion, while the reaction works to homogenize the compositions. This interplay results in a complex steady-state domain morphology, which depends on the reaction rate $\Gamma$ and the relative activity of the species $\Delta$. When $\Delta=1$, we observe only labyrinthine pattern at the steady state, irrespective of the value of $\Gamma$. However, for $\Delta\ne 1$, labyrinthine domains appear for small and intermediate values of $\Gamma$, while the more active component of the mixture forms a droplet-like morphology at sufficiently large values of $\Gamma$ in the steady-state. 

We have characterized the domain growth morphology by computing the equal-time correlation function, $C(r,t)$, and structure factor, $S(k,t)$. The scaled $C(r,t)$s and $S(k,t)$s consistently violate dynamical scaling, irrespective of the values $\Gamma$ and $\Delta$. This violation arises from a structural transition: during the early stages, the system exhibits a diffusion-driven bicontinuous structure~(MIPS), which evolves into a labyrinthine or droplet-like morphology of the more active component of the mixture in the later stages, controlled by the combined effects of diffusion and reaction. However, for $k \rightarrow \infty$, $S(k, t)$ follows Porod's law for all values of $\Gamma$ and $\Delta$. We have estimated the average domain size, $L(t)$, from the decay of $C(r, t)$. At the early stage of evolution, $L(t)$ follows the Lifshitz-Slyozov~(LS) growth law: $L(t) \sim t^{1/3}$. At later times, it approaches a steady-state value, $L_{\rm ss}$. Next, we numerically determined the relationships between $L_{\rm ss}$ and 
$\Gamma$, with $\Delta$ as a parameter. In the steady state, we identified the scaling relation $L_{\rm ss}\sim\Gamma^{-1/4}$, which is independent of $\Delta$.

We believe that our results provide an in-depth understanding of domain kinetics and domain morphology in active binary mixtures in the presence of a chemical reaction. The model we employed can be generalized to multicomponent mixtures with complex interactions among the species, which will be the focus of future studies. Backing our results with agent-based simulations will be the object of a future work. We are aware that discrepancies between our model and agent-based simulations particularly arise at low densities, specifically in the regime where we performed our numerical field-theoretic simulations. However, we are convinced that only quantitative difference will emerge and that the qualitative phenomenology will remain unchanged. Furthermore, our work is relevant to several experimental contexts, spanning biological physics and engineering applications. We hope our findings will inspire further experimental investigations.

\ \\
\noindent{\bf Acknowledgments:} SM acknowledges financial support from IISER Mohali through a Senior Research Fellowship. PD acknowledges financial support from SERB, India through a start-up research grant (SRG/2022/000105).

\ \\
\noindent{\bf Conflict of interest:} The authors have no conflicts to disclose.

\end{document}